\documentclass{elsart}
\usepackage{graphicx}
\begin{document}

\begin{frontmatter}
\title{Quasi-chemical study of Be$^{2+}$(aq) speciation}
\author{D. Asthagiri},
\author{Lawrence R. Pratt\corauthref{cor1}}
\address{Theoretical Division, Los Alamos National Laboratory, Los
Alamos NM 87545} 
\ead{lrp@lanl.gov}
\corauth[cor1]{Corresponding author: Phone: 505-667-8624; Fax: 505-665-3909}
\date{\today}
\begin{abstract}  
Be$^{2+}$(aq) hydrolysis 
can to lead to the formation of multi-beryllium
clusters, but the thermodynamics of this process has not been resolved 
theoretically. We study the hydration state of an isolated
Be$^{2+}$ ion using both the quasi-chemical theory of solutions and 
 ab initio molecular dynamics. These studies confirm that
Be$^{2+}$(aq) is tetra-hydrated. The quasi-chemical approach is then applied to
then  the deprotonation of $\mathrm{Be(H_2O)_4{}^{2+}}$ to
give $\mathrm{BeOH(H_2O)_3{}^{+}}$. The calculated pK$_a$ of 3.8 is in
good agreement with the experimentally suggested value around
3.5. The calculated energetics for the formation of [$\mathrm{Be}
\cdot \mathrm{OH} \cdot \mathrm{Be}$]$^{3+}$ are 
then  obtained in fair agreement with experiments.  
\end{abstract}

\end{frontmatter}

\section{Introduction}

Beryllium metal has  properties that make it
technologically very attractive\cite{vacca:aic00}, but  these
advantages are severely counterbalanced by the high toxicity of
inhaled beryllium dust, which causes chronic beryllium
disease in a subset of exposed individuals \cite{sauer:02}. The
etiology of this autoimmune disease \cite{kotzin:ci01} is
poorly understood, but the final disease state is characterized by
lung failure. 

Aqueous beryllium chemistry is also incompletely understood. Experiments
suggest that beryllium mediated hydrolysis of water leads to the
formation of multi-beryllium clusters
\cite{vacca:aic00,bruno:87}. This same mechanism is likely involved in
the dissolution of Be$^{2+}$(aq) and  of importance in environmental
clean-up strategies. Such mechanisms likely underlie deposition
of beryllium in biomaterials, and thus perhaps in the development  of
chronic beryllium disease.  Thus a molecular  
understanding of Be$^{2+}$(aq) and formation of 
multi-beryllium species would provide a foundation for addressing
these issues of wide importance.  This letter takes an initial
theoretical step  in understanding the aggregation/disaggregation of
beryllium clusters in water.

Early electronic structure calculations on beryllium
hydration were
performed on small clusters 
\cite{iwata:cp87,probst:cpl86} and some attempted to    
include the  second hydration shell in terms of the reaction field
approach \cite{marcos:jpc91} or explicitly \cite{bock:jpc96}.  Molecular dynamics simulation
\cite{probst:cpl89} showed that
assuming pair-wise intermolecular interactions lead a hydration number was six (6) whereas
including three-body effects brought the hydration number down to four
(4) consistent with solution X-ray diffraction experiments.   

Be$^{2+}$ hydration has also been considered within the Car-Parrinello
approach  \cite{Pnello:cpl95,Pnello:ijqc96,Pnello:cpl97}. There a hexa-hydrate cluster quickly dissociated to give a
tetra-hydrate structure \cite{Pnello:cpl95}. The {\em ab initio \/}
molecular dynamics approach for a Be$^{2+}$ atom in a box of 31 water
molecules also showed that the ion was tetra-hydrated
\cite{Pnello:cpl97}. That work also suggested  an
effect of the second hydration shell water molecules on some of 
the bond lengths for the central $\mathrm{Be(H_2O)_4{}^{2+}}$
structure, noted further below. 
\begin{figure}
\begin{center}
\includegraphics{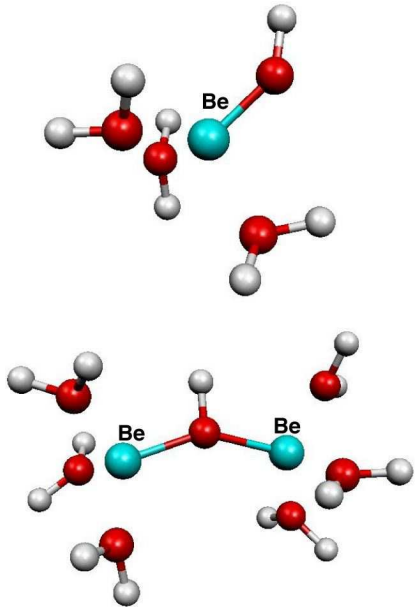}
\end{center}
\caption{Structures representing the deprotonated tetra-aquo cation,
BeOH[H$_2$O]$_3${}$^+$, and the aggregate formed by the coalescence of
one such unit and a Be[H$_2$O]$_4${}$^{2+}$, with
explusion of a water molecule.}\label{fg:cluster} 
\end{figure}

Those earlier works have addressed important issues about Be$^{2+}$
solvation, but the thermodynamical aspect of solvation has not
been considered specifically.  We address the solvation thermodynamics  by
employing two distinct  theoretical approaches. We study the solvation structure of
Be$^{2+}$ using {\em ab initio\/} molecular dynamics methods and confirm
a stable tetra-coordination. The
quasi-chemical theory of solutions then provides a further
analysis why the tetra-hydrate is the most stable species. This
approach has been used before to address  Fe$^{3+}$(aq) speciation
\cite{rlmartin:jpc98}, and the hydration of H$^{+}$ \cite{lrp:jpca02},
HO$^-$ \cite{asthagir:02}, Li$^{+}$ \cite{lrp:jacs00}, and Na$^{+}$
\cite{lrp:fpe01} ions.  As
mentioned above, beryllium solution chemistry holds the
particular challenge of the formation of multi-beryllium clusters. The
present studies on $\mathrm{Be(H_2O)_4{}^{2+}}$ deprotonation and on
formation of [$\mathrm{Be} \cdot \mathrm{OH} \cdot
\mathrm{Be}$]$^{3+}$  provide first steps  in describing that
aggregation process on a molecular  basis, as suggested by
Fig.~\ref{fg:cluster}.

\section{Ab initio molecular dynamics}

The {\em ab initio\/} molecular dynamics (AIMD) simulations were
performed with the  VASP program \cite{kresse:prb93,kresse:prb96}. The
simulation system 
comprises 32 water molecules and one Be$^{2+}$ ion. The box length was
set to 9.71 {\AA} based on the experimental partial molar volume of Be
in water \cite{marcus}.  Vanderbilt ultrasoft pseudopotentials
\cite{vanderbilt,kresse:jpcm94} were used to describe the core-valence
interactions for all the atoms. The valence orbitals were expanded in
plane waves with a kinetic energy cutoff of 29.1 Ry. All the hydrogen atoms
were replaced by deuterium and an integration timestep of 0.5~fs was chosen. 

The initial configuration was obtained by
placing a Be$^{2+}$  ion in  a bath of water molecules.  That
configuration was  energy minimized before
initiating the AIMD simulations. At the end of the classical energy
minimization, the Be$^{2+}$ ion was found to be penta-hydrated
(based on $\mathrm{R(BeO) \leq 2.5}$~{\AA}). In an exploratory 
AIMD run (data not reported), the initial configuration for the AIMD
simulation was obtained from a classical molecular dynamics
simulation. In that case the ion was  hexa-hydrated, but it too
quickly reverted to the tetra-hydrated form.  

In the first 1.0~ps of the AIMD simulation the temperature was
maintained at 300~K by scaling the velocities.  After this initial
phase, a microcanonical (NVE) ensemble simulation was performed for about
2.5~ps.  

Within the first 160~fs of the ca.~3.5~ps of the AIMD simulation, the
coordination number changed to four (4) and stayed so for the rest of
the simulation. The energy in the NVE simulation was
-472.25$\pm$0.08~(2$\sigma$)~eV, suggesting good energy conservation. 
The mean temperature was 316.2$\pm$21.6~K.  

\begin{figure}
\begin{center}
\includegraphics[width=4.25in]{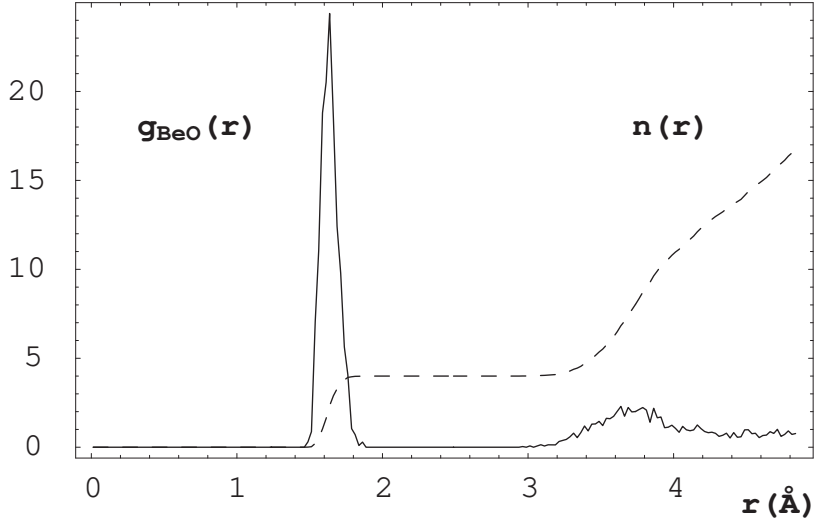}
\end{center}
\caption{Radial distribution of oxygens around Be$^{2+}$. The right
ordinate (dashed line) gives the coordination number, n(r).}\label{fg:gr}
\end{figure}

Fig.~\ref{fg:gr} shows the oxygen(water) radial distribution
around the beryllium ion. Observe that the inner shell is physically
sharply defined and the hydration number is four (4). 
Further structural characterization of Be[H$_2$O]$_4{}^{2+}$ solvation is
collected in table~\ref{tb:aimd}. 
\begin{table}[h]
\caption{Solvation structure of Be[H$_2$O]$_4{}^{2+}$. Inner:
Results for the inner-shell hydration; Outer: Results 
for the outer-shell hydration. Cluster: Results pertaining to a
gas-phase cluster (or isolated molecule) obtained with the methods in
section~III. All lengths are in {\AA} and angles are in
degrees. Values in parentheses are for the  isolated H$_2$O
molecule.}\label{tb:aimd}  
\begin{center}
\begin{tabular}{clcl} \\ \hline
\multicolumn{1}{c}{} & \multicolumn{2}{c}{AIMD} & \multicolumn{1}{c}{} \\
\multicolumn{1}{c}{} & \multicolumn{1}{c}{Inner} &
\multicolumn{1}{c}{Outer} & \multicolumn{1}{c}{Cluster}  \\ \hline 
 R(BeO) &   1.64$\pm$0.06  & --- & 1.64  \\
$\mathrm{\angle{OBeO}}$ & 109.3$\pm$5.9 & --- &   111.4$\pm$5.9\\
ROH  & 1.02$\pm$0.04 & 1.00$\pm$0.03 & 0.98$\pm$0.01 (0.97) \\
$\mathrm{\angle{HOH}} $ & 108.5$\pm$5.6 & 106.5$\pm$5.5 & 111.4 (105.7) \\
\hline \\
\end{tabular}
\end{center}
\end{table}
Earlier\cite{Pnello:cpl97} it was suggested that the second hydration
shell has a significant influence on the OH bond length of the
inner shell water molecule. In the present case, we do find a slight lengthening of the OH bond in
the inner-shell water, but the statistical uncertainties here and in
the earlier study \cite{Pnello:cpl97}, suggest that the difference
between the inner-shell and outer-shell water is subtle. The increase
in the HOH angle for the inner shell water is similar to the earlier
study \cite{Pnello:cpl97} and is in  line with the values obtained 
for an isolated $\mathrm{Be(H_2O)_4{}^{2+}}$ cluster. Also notice that
the oxygen atoms in the inner-shell are nearly tetrahedrally
distributed around the central Be$^{2+}$ ion, but that structure does
fluctuate somewhat.

\section{Quasi-chemical Theory}

In the quasi-chemical theory \cite{lrp:apc02}, the
region around the solute of interest is partitioned into inner and
outer shell domains. In the present study, the inner shell, where
chemical effects are important, is treated quantum mechanically. The
outer shell contributions have been assessed using a dielectric
continuum model. In principle, a variational check of this partition is 
available (see \cite{lrp:fpe01}).

{\bf Hydration state of Be$^{2+}$:} The inner shell reactions
pertinent to Be$^{2+}$ hydration are:
\begin{eqnarray*}
\mathrm{Be^{2+} + n H_2O \rightleftharpoons Be(H_2O)_n{}^{2+}}
\end{eqnarray*}
The free energy change for these reactions were calculated using the
Gaussian programs \cite{gaussian}. The
$\mathrm{Be\cdot(H_2O)_n{}^{2+}}$ (n = 0$\ldots$6) clusters were 
geometry optimized in the gas phase using the B3LYP hybrid density
functional\cite{b3lyp} and the 6-31+G(d,p) basis set. Frequency
calculations confirmed a true minimum, and the zero point energies
were computed at the same level of theory. Single point energies were
calculated using the 6-311+G(2d,p) basis set. A purely inner-shell
$n=5$ cluster could not be obtained; the optimization gave structures
with four (4) inner and one (1) outer sphere water molecule. For $n=6$
both a purely inner-shell configuration, and a structure with four
(4) inner  and two (2) outer shell water molecules were
obtained. The quasi-chemical theory here utilizes only the inner-shell
structure. 

For estimating the outer shell electrostatic contribution, the ChelpG
method \cite{breneman:jcc90} was used to obtain partial atomic
charges. Then with the radii set developed by Stefanovich et
al.\cite{Stefanovich:cpl95}, surface tessera were 
generated \cite{sanner}, and the solvation free energies of the
clusters were calculated using a dielectric continuum model
\cite{lenhoff:jcc90}. With this information and the binding free
energies for the chemical reactions, a primitive quasi-chemical
approximation to the excess chemical potential of Be$^{2+}$(aq) in
water is:
\begin{eqnarray}
\beta \mu_{\mathrm{Be}^{2+}(aq)}^{ex} &\approx& - \ln \left (
\tilde{K}_n \rho_{\mathrm{H}_2\mathrm{O} }{}^n  \right)\label{eq:regrouped}
\end{eqnarray}
where $\tilde{K}_n=K_n{}^{(0)}\exp\left[{-\beta
\left(\mu_{\mathrm{Be}(\mathrm{H}_2\mathrm{O})_n{}^{2+}}^{ex}-n
\mu_{\mathrm{H}_2\mathrm{O} }^{ex}\right)}\right]$. $K_n{}^{(0)}$ is
the equilibrium constant for the reaction in an ideal gas state, with
$n$ of Eq.~\ref{eq:regrouped} the hydration number of the most stable
inner shell cluster, and $\mathrm{\beta=1/k_\mathrm{B}T}$.  The
density factor $\mathrm{\rho_{H_2O}}$ appearing in
eq.~\ref{eq:regrouped} reflects the actual density of liquid water and
its effect is accounted for by including a replacement contribution of
$\mathrm{-n k_\mathrm{B}T \ln (1354)}$.  A detailed statement on standard
states and this replacement contribution can be found in Grabowski et
al.~\cite{lrp:jpca02}.

\begin{figure}
\begin{center}
\includegraphics[width=3.25in]{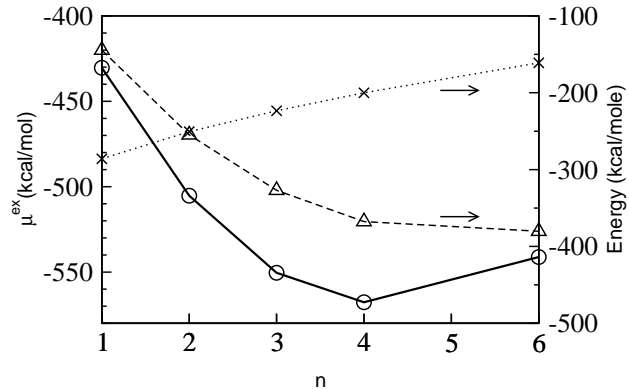}
\end{center}
\caption{Quasi-chemical contributions of the solvation free energy of
Be$^{2+}$(aq). Specifically, the plot (left ordinate) is -k$_B$T
$\times\ln \left 
(\tilde{K}_n \rho_{\mathrm{H}_2\mathrm{O} }{}^n\right)$ {\em vs.\/
}~$n$ predicted by the primitive quasi-chem\-i\-cal approximation; see
Ref.~\cite{lrp:jacs00}.  $n$ is the number of inner shell water
molecules surrounding the anion.  
$\bigtriangleup$: $-RT\ln K_n{}^{0} - n RT\ln(1354)$; 
$\times$:  $\mu_{\mathrm{Be}(\mathrm{H}_2\mathrm{O})_n{}^{2+}}^{ex}-n
\mu_{\mathrm{H}_2\mathrm{O}}$. An observation volume of radius 2.0~{\AA}
centered on the cation defined the inner shell.  Using a smaller radii
did not make an appreciable difference.}\label{fg:beqca}
\end{figure}

From figure~\ref{fg:beqca} it is clear that the tetra-aquo cation is
the most stable form in solution. This is also consistent with the
predictions of the AIMD simulations.  As fig.~\ref{fg:beqca}
indicates, neglecting solvation effects would have forced us to
conclude that both the tetra- and hexa-hydrates 
should be observed. However, it is the substantial unfavorable
solvation of the hexa-hydrate that precludes its presence in the
solvent.  
The calculated solvation free energy of -567.7~kcal/mole is in the
right range  and is in good agreement with the value
(-574.6~kcal/mole) cited in  \cite{marcus}. Solvation free energy 
values for  these highly charged species may have significant
uncertainties, and the agreement here should not mask the real
difficulties facing ion-solvation thermodynamics. The  agreement in
inner-shell hydration number between AIMD simulations and
quasi-chemical calculations is, however, a  
non-trivial  result.

In table~\ref{tb:ener} some of the energies relevant to the $n=4$ case
above is collected in addition to other free energy values that we
will use in analyzing the $\mathrm{pK_a}$ of the tetra-aquo cluster. 
\begin{table}
\caption{Electronic energy (a.u.), thermal corrections (a.u.) to
the free energy, and excess chemical potential (kcal/mole) using
dielectric continuum approximation with charges obtained at
B3LYP/6-311+G(2d,p).}\label{tb:ener} 
\begin{center}
\begin{tabular}{lrrr}\\ \hline
                & \multicolumn{1}{c}{E} &
\multicolumn{1}{c}{G$_{corr}$}  & $\mu^*$ \\ \hline
Be$^{2+}$               & -13.65289     &       -0.01303   & ---   \\
$\mathrm{Be(H_2O)_4{}^{2+}}$  & -320.12296 &  0.07259 & -230.9 \\
$\mathrm{BeOH(H_2O)_3{}^{+}}$ & -319.92375 & 0.05970 & -73.9 \\
$\mathrm{Be_2{}OH(H_2O)_6{}^{3+}}$ & -563.42051 & 0.13225 & -412.0
\\
$\mathrm{Be_2{}OH{}^{3+}}$ & -103.83704 & -0.00627 &  --- \\
H$_2$O                         & -76.45951     &        0.00298   & -7.7  \\
HO$^-$                         &  -75.82779    &       -0.00771   & ---   \\
$\mathrm{HO\cdot(H_2O)_3{}^-}$ & -305.32036    &        0.04705   &
-72.3 \\ \hline \\
\end{tabular}
\end{center}
\end{table}

{\bf $\mathrm{pK_a}$ of $\mathrm{Be(H_2O)_4{}^{2+}}$:} The acidity of $\mathrm{Be(H_2O)_4{}^{2+}}$ is
described by the 
\begin{eqnarray}
K_a = \frac{\left\lbrack\mathrm{BeOH(H_2O)_3{}^{+}}\right\rbrack\left\lbrack\mathrm{H{}^{+}}\right\rbrack}{\left\lbrack\mathrm{Be(H_2O)_4{}^{2+}}\right\rbrack}
\end{eqnarray}
corresponding to the reaction\begin{eqnarray}
\mathrm{Be(H_2O)_4{}^{2+}} \rightleftharpoons \mathrm{BeOH(H_2O)_3{}^{+} + H^+}
\label{eq:disso1}
\end{eqnarray}
under standard conditions (1~M ideal solution).  The
reaction as written requires us to know the hydration free energy of
the proton, estimates of which have large uncertainties
\cite{lrp:jpca02}. Alternatively, using a reference reaction such as the 
dissociation of water can obviate the need for knowing the proton
hydration free energy. Thus consider
\begin{eqnarray}
\mathrm{Be(H_2O)_4{}^{2+} + HO^-} \rightleftharpoons
\mathrm{BeOH(H_2O)_3{}^{+} + H_2O}
\label{eq:disso2}
\end{eqnarray}
Knowing the energetics of eq.~\ref{eq:disso2}  permits calculation of the $\mathrm{pK_a}$ of the
tetra-aquo cation, according to
\begin{eqnarray}
K_a = K \times \left(K_w/\left\lbrack\mathrm{H_2O}\right\rbrack\right)~.
\label{kw}
\end{eqnarray}  
Here $K$ is the  equilibrium ratio of eq.~\ref{eq:disso2} with all concentration units the same, $K_w$ = $\mathrm{\left\lbrack HO^-\right\rbrack\left\lbrack H^+\right\rbrack}$ the standard ion product for water; here
$\mathrm{pK_w}=15.7$ \cite{pearson:jacs86}.  An added advantage of using a reference reaction
as above is that some  cancellation of errors can be encouraged.

The excess chemical potential of HO$^-$(aq) in
reaction~\ref{eq:disso2} above is also obtained within the
quasi-chemical approach and with the tri-hydrated quasi-component
\cite{asthagir:02}. The computed value of -105~kcal/mole is in good
agreement with recently reported values \cite{tuttle:jpca98}. 

If all standard concentrations are 1~M, then using the values in table~\ref{tb:ener} the free energy change for
reaction~\ref{eq:disso2} is -18.5~kcal/mole. Eq.~\ref{kw} then yields $\mathrm{pK_a}\approx$ 3.8.    

By fitting experimental free energy changes for the case of
low total Be$^{2+}$ concentration \cite{vacca:aic00,bruno:87}, it is 
found that $\mathrm{Be(H_2O)_4{}^{2+}}$ exists
in appreciable amounts only below a pH of 3.5. The present calculated
$\mathrm{pK_a}$ is in excellant agreement with these observations. This value of
$\mathrm{pK_a}$ has the standard interpretation that the deprotonated
complex $\mathrm{BeOH(H_2O)_3{}^{+}}$ is  above a thousand times more probable
than $\mathrm{Be(H_2O)_4{}^{2+}}$ at neutral pH.
That a spontaneous deprotonation is not observed in our simulations is
reflective of the limited simulation time and  the possibility
of this process being activated.   Nevertheless, when an OH$^-$ was introduced
into the AIMD simulation by extraction of a proton distant from the beryllium ion,
ligand exchange by proton shuffling was accomplished in less than a ps.

{\bf Formation of [$\mathrm{Be} \cdot \mathrm{OH} \cdot
\mathrm{Be}$]$^{3+}$:} The [$\mathrm{Be} \cdot \mathrm{OH} \cdot
\mathrm{Be}$]$^{3+}$ cluster is one of the many clusters that
beryllium forms \cite{vacca:aic00}, but it is the simplest. 
Other clusters could be constructed with this unit. Thus understanding
the formation 
thermodynamics of this cluster is of first interest. 

The solvation free energy of the complex [$\mathrm{Be} \cdot
\mathrm{OH} \cdot \mathrm{Be}$]$^{3+}$ is obtained from the following
reaction.  
\begin{eqnarray}
\mathrm{Be\cdot OH\cdot Be^{3+} + 6 H_2O \rightleftharpoons 
Be\cdot OH\cdot Be \cdot(H_2O)_6{}^{3+}}
\end{eqnarray}
Here is is assumed that each of the originally tetra-hydrated
Be$^{2+}$ loses one water and gains a HO$^-$ to form the
complex (fig.~\ref{fg:cluster}). Using eq.~\ref{eq:regrouped} for the
present case, the 
quasi-chemical estimate of the solvation free energy of 
[$\mathrm{Be} \cdot \mathrm{OH} \cdot \mathrm{Be}$]$^{3+}$ is obtained
as -834.5~kcal/mole.

With this solvation free energy estimate, we can enquire about the
thermodynamics of the complexation reaction below.
\begin{eqnarray}
\mathrm{ 2 Be^{2+} + HO^- \rightleftharpoons [Be\cdot OH\cdot Be]^{3+}
}
\end{eqnarray}
The change in the \emph{excess} chemical potential for
the reaction  is 405.9~kcal/mole, whereas the \emph{ideal} contribution
{\em  i.e.,\/} for an
the ideal gas at 1~atm pressure, is -424.2~kcal/mole for this change. This gives
the \emph{net} free energy change of -18.3~kcal/mole not accounting for standard 
concentration. Converting to  the  standard concentration of 1~M
adjusts this by $\mathrm{-2\cdot RT \ln 24.46
}$ to  the calculated free energy change of -22.1~kcal/mole.  

An experimental value for the complexation reaction is
-14.4~kcal/mole\cite{vacca:aic00,bruno:87}. The calculated energetics
are in the right range, but the agreement is only fair. Note also that
a small difference between large numbers is being computed; thus even
minor errors will tend to get amplified. A physical conclusion is that the hydration
contribution is more 20 times larger than the net standard free energy change for this  
reaction; neglecting hydration effects would lead to a qualitatively incorrect result.
Below we consider ways to improve upon these initial estimates.  

\section{Concluding Discussions}

The quasi-chemical approach leads to free energies that are
in reasonable agreement with available experimental estimates. But this
agreement should not obscure the severe approximations that have been
made in applying the theory to practical calculations. For divalent
cations, it is unclear if purely a inner-shell complex would provide
adequate accuracy of solvation free energy estimate. There are
potentially two possible avenues for  improvement which we indicate below.  

First, the outer-sphere hydration contributions can be
obtained  using classical molecular mechanics approaches instead of
the dielectric continuum model used above. Such an approach is now
being undertaken for water clusters on monovalent cations (Asthagiri, et
al. in preparation)  and this can likely be used for
$\mathrm{Be(H_2O)_4{}^{2+}}$ as well.  

Second, it is possible to expand the chemical potential of the
tetra-aqua cation in terms of its own inner-shell (i.e. the Be$^{2+}$
cation's 2$^{nd}$ shell). This seems particularly natural here because
the inner shell structures are physically definite
(fig.~\ref{fg:gr}). The hydration contribution of those 2$^{nd}$
shell structures could be obtained using a dielectric continuum
model. Also it is possible to apply a lower level of quantum chemical
approximation to describe the coupling of the 1$^{st}$ and 2$^{nd}$
shells.  Then with an improved estimate for the solvation free  
energy of the tetra-aqua cation, one could obtain a better estimate
for the solvation free energy of Be$^{2+}$. This approach is the
iterated quasi-chemical scheme and a variant has been successfully
applied to  hard-sphere fluids \cite{lrp:jpcb02}. In the present case,
however, this approach would be daunting for many reasons. (A)
Obtaining statistically representative 2$^{nd}$-shell complexes in the
gas-phase is difficult. (B) The rigid-rotor harmonic oscillator
approximation is dubious for applications to such clusters. (C)
Applying the  quasi-chemical theory to such large clusters demands
consideration of packing effects. This latter issue is actively under
study at present (Ashbaugh and Pratt, in preparation). 

The neglect of packing aspect is certainly one the reason underlying
the merely fair agreement  for the [$\mathrm{Be} \cdot \mathrm{OH}
\cdot \mathrm{Be}$]$^{3+}$ thermochemistry. It is also likely
that anharmonic effects   are non-negligible for clusters such as
$\mathrm{Be_2{}OH^{3+}}$ and $\mathrm{Be_2{}OH(H_2O)_6{}^{3+}}$.
But despite these substantial limitations, it is  heartening to obtain
the qualitatively correct trends. In our $\mathrm{pK_a}$ calculations,
however, by arranging the equation  to have similarly sized species on
both sides of the equality we were able to mitigate these
uncertainties.

\section{Acknowledgements}
The work at Los Alamos was supported by the US Department of Energy,
contract W-7405-ENG-36, under the LDRD program at Los Alamos. LA-UR-03-0073.


\end{document}